# On spin evolution in a time-dependent magnetic field: post-adiabatic corrections and geometric phases

K.Yu. Bliokh[a,b,c,*]

[a]*Institute of Radio Astronomy, 4 Krasnoznamyonnaya St., Kharkov, 61002, Ukraine*
[b]*Amcrys Ltd., 60 Lenin Ave., Kharkov, 61001, Ukraine*
[c]*Optical Engineering Laboratory, Faculty of Mechanical Engineering,
Technion–Israel Institute of Technology, Haifa 32000, Israel*

We examine both quantum and classical versions of the problem of spin evolution in a slowly varying magnetic field. Main attention is given to the first- and second-order adiabatic corrections in the case of in-plane variations of the magnetic field. While the first-order correction relates to the classical Berry phase and Coriolis-type lateral deflection of the spin, the second-order correction is shown to be responsible for the next-order geometric phase and in-plain deflection. A comparison between different approaches, including the exact (non-adiabatic) geometric phase, is presented.

## 1. INTRODUCTION

The problem of the spin evolution in a time-dependent magnetic field is of a great importance both in classical physics and quantum mechanics. In the quantum case, for spin 1/2, it represents the basic model of a driven two-level system with $SU(2)$ evolution, whereas the classical (mean) spin vector obeys the Bloch precession equation [1] with $SO(3)$ evolution which can be represented on $S^2$ sphere. In the constant field, quantum eigenstates correspond to the stationary solutions for the classical spin vector, and the phase difference between the eigenstates is responsible for the precession frequency of other solutions. If the magnetic field varies slowly as compared to the eigenfrequencies of the system, one can make use of the adiabatic approximation. In the zero-order approximation, the classical spin locally precesses around the instant direction of the magnetic field. In this approximation, the classical quasi-stationary spin states are aligned with the magnetic field, and the quantum phase of adiabatic eigenstate is the usual dynamical phase.

In the first approximation, the classical quasi-stationary spin states undergo a small deflection due to the Coriolis-type force [2], which corresponds to a small deformation of the eigenstates polarization. At the same time, as was shown by Berry, the quantum phase acquires an additional term: the adiabatic geometric phase [3]. The Berry phase appears due to the presence of natural $U(1)$ 'magnetic monopole' connection in the principal fiber bundle over the parameter unit sphere, associated with the direction of the magnetic field, which provides for the parallel transport of the quantum adiabatic state vector. For cyclic evolutions, the Berry phase is numerically equal (up to 1/2 spin factor) to the solid angle enclosed by the loop on the magnetic-field sphere [3]. A classical counterpart of the Berry phase is an additional angle in the classical spin precession, which shifts the average precession frequency. This is the Hannay angle similar to that in example of the rotated rotator [4].

The adiabatic consideration of the problem is usually restricted within the first approximation. However, higher-order corrections become essential at long enough times (since they contribute to effective eigenfrequencies [9]), especially if the first-order correction vanishes. The explicit calculation of the higher-order corrections is also important for the validation of the adiabatic approximation, which has been doubted recently [5].

---

[*]E-mail: k_bliokh@mail.ru

The higher-order adiabatic description of the spin evolution in a magnetic field was investigated by several authors and has led to three different approaches to the problem. First, Berry proposed a solution to the problem with cyclic evolution by constructing the iteration series of unitary transformations of the original time-dependent Hamiltonian [6]. In so doing, he found a $\mathbb{R}^3 \to \mathbb{R}^3$ map, which enables one to find the adiabatic solution up to any order. In Berry's method the phase corrections contain terms of two different origins: corrections to the eigenfrequencies, (they are even-order) and corrections to the geometric phase (they are odd-order), which are described by the same 'magnetic monopole' connection in the fiber bundle over the effective magnetic field unit sphere.

Second, there is an exact, non-adiabatic expression for the quantum phase, which is also known as the Aharonov−Anandan phase [7]. For the system under consideration, the exact phase is determined by the same simple geometrical law ('magnetic monopole' connection) on the sphere, but, this time, this is the classical-spin sphere rather than the magnetic-field parameter one. In other words, *the exact quantum phase is expressed in terms of an unknown solution of the classical equation of motion*, i.e. the Bloch equation in our case. This is a consequence of the fact that quantum and classical spin evolutions are related through the Hopf fibration $\mathrm{SU}(2) \to \mathrm{S}^2 \cong \mathrm{SU}(2)/\mathrm{U}(1)$ yielding $\mathrm{U}(1)$ principal fiber bundle over the classical spin sphere. It is worth remarking, that this correspondence between the quantum phase and classical spin evolution has been known before the Berry phase discovery in the context of the semiclassical evolution of spinning particles (see, e.g., [8]). Thus, all the phase corrections discussed by Berry [6] can be represented in the form of the single geometrical phase [7,8].

Finally, a general method to construct adiabatic solutions up to any power of the small parameter of adiabaticity for time-dependent Hamiltonians was proposed in [9]. It is similar to the method of [6] and consists in the recursive diagonalizations of the initial Hamiltonian (which is assumed to be non-degenerated). All the phase corrections have been separated there into the local (integrable) and non-local (non-integrable) terms. The latter were represented in the form of the geometric phase on the generalized parameter space which included the time derivatives of the parameters as independent dimensions. This formal approach has been successfully applied to concrete physical systems [10]. As shown in [9], all the non-local corrections from the adiabatic series in the phase can be considered both as generalized geometrical phases and as corrections to the eigenfrequencies of the system.

In the present paper we analyze quantum and classical versions of the spin evolution problem in the second-order adiabatic approximation. Basically, we will consider in-plane variations of the magnetic field, because this is the case when the first-order Berry phase vanishes. (For simplicity, we do not consider the complete rotations of the magnetic field resulting in the $-\pi$ Berry phase.) We will present quantum and classical solutions, analyze the phase, and compare all the above-mentioned approaches. Note that a related comparison between adiabatic and exact geometrical phases has been given recently in [11] for the case of a uniformly rotating magnetic field.

## 2. THE SECOND-ORDER SOLUTION IN SPINOR REPRESENTATION

The problem of the evolution of spin 1/2 in a slowly-varying in-plane magnetic field $\mathbf{B} = \mathbf{B}(\varepsilon t)$ is described by Hamiltonian ($\hbar = 1$)

$$\hat{H} = \mu \mathbf{B} \frac{\hat{\boldsymbol{\sigma}}}{2}, \qquad (1)$$

where $\hat{\boldsymbol{\sigma}}$ is the vector of Pauli matrices, $\mu$ is a constant involving gyromagnetic ratio, and $\varepsilon \ll 1$ is the small parameter of adiabaticity. Here and in what follows all matrices are denoted with hats. To eliminate the constant $\mu$ from the consideration we will use rescaled energy- or frequency-dimension field $\mathcal{B} = \mu \mathbf{B}$. We will consider formal asymptotic in $\varepsilon$, however the actual dimensionless small parameter of the adiabaticity is $\varepsilon / \mathcal{B}$. For the field evolving in $(x, z)$ plane we have $\mathcal{B} \equiv (\mathcal{B}_x, \mathcal{B}_y, \mathcal{B}_z) = \mathcal{B}(\sin\theta, 0, \cos\theta)$, where $\theta$ is the polar angle. Then, the Schrödinger equation and Hamiltonian (1) read

$$i|\dot{\psi}\rangle = \hat{H}(\varepsilon t)|\psi\rangle, \quad \hat{H} = \mathcal{B}\frac{\hat{\boldsymbol{\sigma}}}{2} = \frac{\mathcal{B}}{2}\begin{pmatrix} \cos\theta & \sin\theta \\ \sin\theta & -\cos\theta \end{pmatrix}. \tag{2}$$

Here $|\psi\rangle = \begin{pmatrix} \psi^\uparrow \\ \psi^\downarrow \end{pmatrix}$ is the two-component spinor wave function and the overdot stands for the time derivative.

Equation (2) can be solved through recursive diagonalizations [9]. There is a convenient U(2) iteration procedure proposed in [12] for Eq. (2). Nonetheless, we will apply slightly different SU(2) diagonalizations, restricting ourselves to an accuracy of $\varepsilon^2$ in calculations. A unitary transformation $|\psi_j\rangle = \hat{U}_j(\varepsilon t)|\psi_{j+1}\rangle$ in the Schrödinger equation $i|\dot{\psi}_j\rangle = \hat{H}_j(\varepsilon t)|\psi_j\rangle$ leads to the transformation in the Hamiltonian, $\hat{H}_{j+1} = \hat{U}_j^\dagger \hat{H}_j \hat{U}_j - i\hat{U}_j^\dagger \dot{\hat{U}}_j$, and the next-step Schrödinger equation $i|\dot{\psi}_{j+1}\rangle = \hat{H}_{j+1}(\varepsilon t)|\psi_{j+1}\rangle$. Applying three successive diagonalizations (so that $\hat{U}_j^\dagger \hat{H}_j \hat{U}_j$ is a diagonal matrix at each step) to the equation (2), we have ($|\psi\rangle \equiv |\psi_0\rangle$, $\hat{H} \equiv \hat{H}_0$)

$$\hat{U}_0 = \begin{pmatrix} \cos\frac{\theta}{2} & -\sin\frac{\theta}{2} \\ \sin\frac{\theta}{2} & \cos\frac{\theta}{2} \end{pmatrix}, \quad \hat{H}_1 = \frac{\mathcal{B}}{2}\begin{pmatrix} 1 & i\delta \\ -i\delta & -1 \end{pmatrix}; \tag{3}$$

$$\hat{U}_1 = \begin{pmatrix} 1-\frac{\delta^2}{8} & -i\frac{\delta}{2} \\ -i\frac{\delta}{2} & 1-\frac{\delta^2}{8} \end{pmatrix} + O(\varepsilon^3), \quad \hat{H}_2 = \frac{\mathcal{B}^{eff}}{2}\begin{pmatrix} 1 & -\gamma \\ -\gamma & -1 \end{pmatrix} + O(\varepsilon^3); \tag{4}$$

$$\hat{U}_2 = \begin{pmatrix} 1 & \frac{\gamma}{2} \\ -\frac{\gamma}{2} & 1 \end{pmatrix} + O(\varepsilon^3), \quad \hat{H}_3 = \frac{\mathcal{B}^{eff}}{2}\text{diag}(1,-1) + O(\varepsilon^3). \tag{5}$$

Here

$$\delta = \frac{\dot{\theta}}{\mathcal{B}} \sim \varepsilon \quad \text{and} \quad \gamma = \frac{\dot{\delta}}{\mathcal{B}} = \frac{\ddot{\theta} - \dot{\theta}\dot{\mathcal{B}}/\mathcal{B}}{\mathcal{B}^2} \sim \varepsilon^2 \tag{6}$$

are the parameters related to the angular velocity and acceleration of the magnetic field evolution, whereas

$$\mathcal{B}^{eff} = \mathcal{B}\left(1 + \frac{\delta^2}{2}\right) \tag{7}$$

is the effective field strength determining the slightly shifted (on the order of $\varepsilon^2$) energy levels of the system. The geometrical meaning of diagonalizations (3)–(5) will be revealed in the next Section.

Equations (6) and (7) show that the shift of eigenfrequencies relative to the instant values $\pm \mathcal{B}/2$ is proportional to the square of angular velocity of the magnetic field rotation, $\dot{\theta}^2$, and is independent of the velocity of the magnetic field strength variations, $\dot{\mathcal{B}}$. As it follows from [12], the eigenfrequencies acquire only even-order corrections, which is directly related to the vanishing of the Berry phase for in-plane magnetic field evolution. (In the general case, the Berry phase results in the $\varepsilon$-order correction [9].) At the same time, the spinor polarization and the direction of the classical spin vector contains corrections of all the orders.

Taking this into account, we can write a general solution of the Schrödinger equation with Hamiltonian (5) as:

$$|\psi_3\rangle = \begin{pmatrix} \alpha e^{i\varphi} \\ \beta e^{-i\varphi} \end{pmatrix} + O(\varepsilon^3, \varepsilon^4 t), \tag{8}$$

where $\alpha$ and $\beta$ are complex constants with normalization $|\alpha|^2 + |\beta|^2 = 1$, whereas the phase $\varphi$ equals

$$\varphi = -\frac{1}{2}\int_0^t \mathcal{B}^{eff} dt' \equiv \varphi^{(0)} + \varphi^{(2)}, \quad \varphi^{(0)} = -\frac{1}{2}\int_0^t \mathcal{B}\, dt', \quad \varphi^{(2)} = \int_0^t \frac{\dot{\theta}^2}{4\mathcal{B}} dt', \tag{9}$$

where $\varphi^{(0)} \sim \varepsilon^0$ is the usual dynamical phase and $\varphi^{(2)} \sim \varepsilon^2$ is the second-order correction. Although the phase (9) is represented as a dynamical one from $\mathcal{B}^{eff}$ magnetic field, in Section 4 we will show that $\varphi^{(2)}$ can be associated with the second-order geometric phase correction. The phase $\varphi^{(2)}$ is non-local; it can grow unlimitedly with time and can be estimated by the order of magnitude as $\varphi^{(2)} \sim \varepsilon^2 t / \mathcal{B}$. Hence at any finite $\varepsilon$ there exists a time $t_1 \sim \mathcal{B}/\varepsilon^2$, when phase $\varphi^{(2)}$ becomes significant. Note that solutions (8) and (9) are applicable for the times much smaller than $t_2 \sim \mathcal{B}^3 / \varepsilon^4 \gg t_1$, where the next, $\varepsilon^4$-order, correction to the frequency become noticeable.

Solution (8) can be transformed back to the original basis as $|\psi\rangle = \hat{U}_0 \hat{U}_1 \hat{U}_2 |\psi_3\rangle$. Calculations with Eqs. (3)–(5) result in

$$|\psi\rangle = \alpha e^{i\varphi} \begin{pmatrix} \left(1 - \frac{\delta^2}{8}\right)\cos\frac{\theta}{2} + i\frac{\delta}{2}\sin\frac{\theta}{2} + \frac{\gamma}{2}\sin\frac{\theta}{2} \\ \left(1 - \frac{\delta^2}{8}\right)\sin\frac{\theta}{2} - i\frac{\delta}{2}\cos\frac{\theta}{2} - \frac{\gamma}{2}\cos\frac{\theta}{2} \end{pmatrix}$$

$$+ \beta e^{-i\varphi} \begin{pmatrix} -\left(1 - \frac{\delta^2}{8}\right)\sin\frac{\theta}{2} - i\frac{\delta}{2}\cos\frac{\theta}{2} + \frac{\gamma}{2}\cos\frac{\theta}{2} \\ \left(1 - \frac{\delta^2}{8}\right)\cos\frac{\theta}{2} - i\frac{\delta}{2}\sin\frac{\theta}{2} + \frac{\gamma}{2}\sin\frac{\theta}{2} \end{pmatrix} + O(\varepsilon^3, \varepsilon^4 t) \tag{10}$$

## 3. THE SECOND-ORDER SOLUTION ON CLASSICAL SPIN SPHERE

Let us examine now the evolution of the mean (classical) unit spin vector

$$\mathbf{S} = \langle \psi | \boldsymbol{\sigma} | \psi \rangle, \tag{11}$$

or

$$S_x = 2\operatorname{Re}\left(\psi^{\uparrow *}\psi^{\downarrow}\right), \quad S_y = 2\operatorname{Im}\left(\psi^{\uparrow *}\psi^{\downarrow}\right), \quad S_z = |\psi^{\uparrow}|^2 - |\psi^{\downarrow}|^2. \tag{11'}$$

By differentiating definition (11) and using Schrödinger equation (2) along with the commutation relations for Pauli matrices, $\left[\hat{\sigma}_i, \hat{\sigma}_j\right] = 2ie_{ijk}\hat{\sigma}_k$, we arrive at the Bloch equation for the classical spin precession:

$$\dot{\mathbf{S}} = \mathcal{B} \times \mathbf{S}. \qquad (12)$$

This equation is just a consequence of the Ehrenfest theorem, since in the Heisenberg representation the equation of motion for the spin operator, $\hat{\boldsymbol{\sigma}}^H$, reads $\dot{\hat{\boldsymbol{\sigma}}}^H = i\left[\hat{H}, \hat{\boldsymbol{\sigma}}\right]^H = \mathcal{B} \times \hat{\boldsymbol{\sigma}}^H$.

The successive complex diagonalizations can also be applied to the equation (12) given in the matrix form. However, it is more convenient and meaningful to use successive SO(3) rotation transformations. Indeed, the precession equation (12) is easy to solve if the magnetic field is directed along, e.g., $z$ axis. Hence, our aim is to superpose the $z$ axis with the instant $\mathcal{B}$ direction. Since $\mathcal{B}$ lies in $(x, z)$ this is realized by the rotation at angle $\theta$ about $y$ axis leading to the transformation

$$\mathbf{S} = \hat{R}_0(\varepsilon t)\mathbf{S}_1, \quad \hat{R}_0 = \begin{pmatrix} \cos\theta & 0 & \sin\theta \\ 0 & 1 & 0 \\ -\sin\theta & 0 & \cos\theta \end{pmatrix}. \qquad (13)$$

Such transformation for Eq. (12) yields the equation $\dot{\mathbf{S}}_1 = \hat{R}_0^{-1}\mathcal{B} \times \hat{R}_0\mathbf{S}_1 - \hat{R}_0^{-1}\dot{\hat{R}}_0$, which can be written as

$$\dot{\mathbf{S}}_1 = \mathcal{B}_1 \times \mathbf{S}_1, \quad \mathcal{B}_1 = \mathcal{B}(0, -\delta, 1). \qquad (14)$$

Thus, the rotation (13) induces a small $y$ component of the effective magnetic field, responsible for the Coriolis force [2].

To superpose the $z$ axis with the instant $\mathcal{B}_1$ direction, Eq. (14), we should turn the coordinate frame at small angle $\delta \sim \varepsilon$ ($\sin\delta \simeq \delta$ and $\cos\delta \simeq 1 - \delta^2/2$) about $x$ axis. In so doing, we have the transformation in Eq. (14) as follows:

$$\mathbf{S}_1 = \hat{R}_1 \mathbf{S}_2, \quad \hat{R}_1 = \begin{pmatrix} 1 & 0 & 0 \\ 0 & 1 - \dfrac{\delta^2}{2} & -\delta \\ 0 & \delta & 1 - \dfrac{\delta^2}{2} \end{pmatrix} + O(\varepsilon^3), \qquad (15)$$

which results in

$$\dot{\mathbf{S}}_2 = \mathcal{B}_2 \times \mathbf{S}_2, \quad \mathcal{B}_2 = \mathcal{B}^{eff}(-\gamma, 0, 1) + O(\varepsilon^3), \qquad (16)$$

where the small $x$-component of the field $\mathcal{B}_2$ appeared because of the second-order Coriolis term caused by small rotation (15). The final step is a rotation at the angle $-\gamma \sim \varepsilon^2$ ($\sin\gamma \simeq \gamma$ and $\cos\gamma \simeq 1$) about $y$ axis:

$$\mathbf{S}_2 = \hat{R}_2 \mathbf{S}_3, \quad \hat{R}_3 = \begin{pmatrix} 1 & 0 & -\gamma \\ 0 & 1 & 0 \\ \gamma & 0 & 1 \end{pmatrix} + O(\varepsilon^3), \qquad (17)$$

yielding the equation

$$\dot{\mathbf{S}}_3 = \mathcal{B}_3 \times \mathbf{S}_3, \quad \mathcal{B}_3 = \mathcal{B}^{eff}(0, 0, 1) + O(\varepsilon^3). \qquad (18)$$

Thus, we arrive at the desired precession equation with the effective magnetic field directed along $z$ axis in the approximation under consideration. The general solution of (18) is

$$\mathbf{S}_3 = \begin{pmatrix} A\cos 2\varphi - B\sin 2\varphi \\ A\sin 2\varphi + B\cos 2\varphi \\ C \end{pmatrix} + O(\varepsilon^3, \varepsilon^4 t), \qquad (19)$$

where $A$, $B$, and $C$ are real constants with normalization $A^2 + B^2 + C^2 = 1$, and $\varphi$ is given by Eq. (9). Equation (19) describes the spin precession around the effective magnetic field $\mathcal{B}_3$. There, $2\varphi$ is the azimuthal angle of classical spin in the exploited coordinate frame, so that the second-order correction $\varphi^{(2)}$ in Eq. (9) corresponds to the second-order Hannay angle $2\varphi^{(2)}$ [4] and indicates the frequency shift for the classical spin precession, $\mathcal{B} \to \mathcal{B}^{\text{eff}}$, Eq. (7). Solutions (19) with $C = \pm 1$ correspond to the quasi-stationary states without precession, when the spin is parallel or antiparallel to $\mathcal{B}_3$. These are the higher- and lower-energy states, corresponding to the quantum states with $\alpha = 1$ or $\beta = 1$ in Eqs. (8) and (10). Solution (19) can be transformed back to the original coordinate frame of Eq. (12): $\mathbf{S} = \hat{R}_0 \hat{R}_1 \hat{R}_2 \mathbf{S}_3$. Calculating with Eqs. (13), (15), and (17), we obtain

$$\mathbf{S} = \begin{pmatrix} A(\cos\theta + \gamma\sin\theta) + B\delta\sin\theta \\ B\left(1 - \dfrac{\delta^2}{2}\right) \\ A(-\sin\theta + \gamma\cos\theta) + B\delta\cos\theta \end{pmatrix} \cos 2\varphi + \begin{pmatrix} -B(\cos\theta + \gamma\sin\theta) + A\delta\sin\theta \\ A\left(1 - \dfrac{\delta^2}{2}\right) \\ B(\sin\theta - \gamma\cos\theta) + A\delta\cos\theta \end{pmatrix} \sin 2\varphi$$

$$+ C \begin{pmatrix} \left(1 - \dfrac{\delta^2}{2}\right)\sin\theta - \gamma\cos\theta \\ -\delta \\ \left(1 - \dfrac{\delta^2}{2}\right)\cos\theta + \gamma\sin\theta \end{pmatrix}, \qquad (20)$$

with $O(\varepsilon^3, \varepsilon^4 t)$ omitted in what follows.

By substituting spinor solution (10) into (11') one can make sure explicitly that it leads to the classical solution (20). In this way, we find that

$$A = 2\operatorname{Re}(\alpha^* \beta), \quad B = 2\operatorname{Im}(\alpha^* \beta), \quad C = |\alpha|^2 - |\beta|^2. \qquad (21)$$

This implies that $\mathbf{S}_3 = \langle \psi_3 | \boldsymbol{\sigma} | \psi_3 \rangle$, i.e. SU(2) diagonalization transformations of the previous Section exactly correspond to the geometric SO(3) rotations presented in this Section.

Let us analyse qusi-stationary solution (20), $A = B = 0$ and $C = 1$, in detail:

$$\mathbf{S} = \begin{pmatrix} \left(1 - \dfrac{\delta^2}{2}\right)\sin\theta - \gamma\cos\theta \\ -\delta \\ \left(1 - \dfrac{\delta^2}{2}\right)\cos\theta + \gamma\sin\theta \end{pmatrix} \equiv \mathbf{S}^{(0)} + \mathbf{S}^{(1)} + \mathbf{S}^{(2)}. \qquad (22)$$

where

$$\mathbf{S}^{(0)} = \begin{pmatrix} \sin\theta \\ 0 \\ \cos\theta \end{pmatrix}, \quad \mathbf{S}^{(1)} = -\delta\begin{pmatrix} 0 \\ 1 \\ 0 \end{pmatrix}, \quad \mathbf{S}^{(2)} = \begin{pmatrix} -\gamma\cos\theta - \dfrac{\delta^2}{2}\sin\theta \\ 0 \\ \gamma\sin\theta - \dfrac{\delta^2}{2}\cos\theta \end{pmatrix}. \qquad (23)$$

Here $\mathbf{S}^{(j)} \sim \varepsilon^j$, i.e. $\mathbf{S}^{(0)}$ is the zero-order adiabatic solution, while $\mathbf{S}^{(1)}$ and $\mathbf{S}^{(2)}$ are the first- and second- order adiabatic corrections to it. Solution (23) can also be given as

$$\mathbf{S}^{(0)} = \frac{\mathcal{B}}{\mathcal{B}}, \quad \mathbf{S}^{(1)} = \frac{\dot{\mathbf{S}}^{(0)} \times \mathbf{S}^{(0)}}{\mathcal{B}}, \quad \mathbf{S}^{(2)} = \frac{\dot{\mathbf{S}}^{(1)} \times \mathbf{S}^{(0)}}{\mathcal{B}} - \frac{\left|\mathbf{S}^{(1)}\right|^2}{2} \mathbf{S}^{(0)}. \tag{24}$$

By the direct substitution one can ascertain that solution (24) satisfies (12) in the general case of *3-dimensional* adiabatic evolution of the magnetic field. Equation (24) also reveals the relation of the adiabatic corrections to the inertia forces. The first correction is the usual Coriolis force [2] accompanying the Berry phase in 3D case [3,6], while the second-order correction is the second-order Coriolis force. The $\mathbf{S}^{(0)}$-directed term in $\mathbf{S}^{(2)}$ appears only to provide the unitarity of $\mathbf{S}$: $\left|\mathbf{S}^{(0)} + \mathbf{S}^{(1)} + \mathbf{S}^{(2)}\right|^2 = 1 + O(\varepsilon^3)$. In the case of in-plane magnetic field evolution, assuming $\dot{\mathcal{B}} = 0$, we can present solution (23) with Eq. (6) in spherical coordinate system $(r, \theta, \phi)$ as

$$\mathbf{S}^{(0)} = \mathbf{e}_r, \quad \mathbf{S}^{(1)} = -\frac{\dot{\theta}}{\mathcal{B}} \mathbf{e}_\phi, \quad \mathbf{S}^{(2)} = -\frac{\ddot{\theta}}{\mathcal{B}^2} \mathbf{e}_\theta, \tag{25}$$

where $\mathbf{e}_k$ are the respective unit vectors. $\mathbf{S}^{(1)}$ is equal in absolute value to dimensionless velocity $\dot{\theta}/\mathcal{B}$ of the motion on the sphere and is directed orthogonally to it, whereas $\mathbf{S}^{(2)}$ is equal to dimensionless acceleration $\ddot{\theta}/\mathcal{B}^2$ and directed oppositely to it. Hence, the second-order correction can also be associated with the usual inertia related to the linear accelerated motion on the sphere. Figure 1 shows the directions of the adiabatic corrections (24) or (25) on the classical-spin sphere.

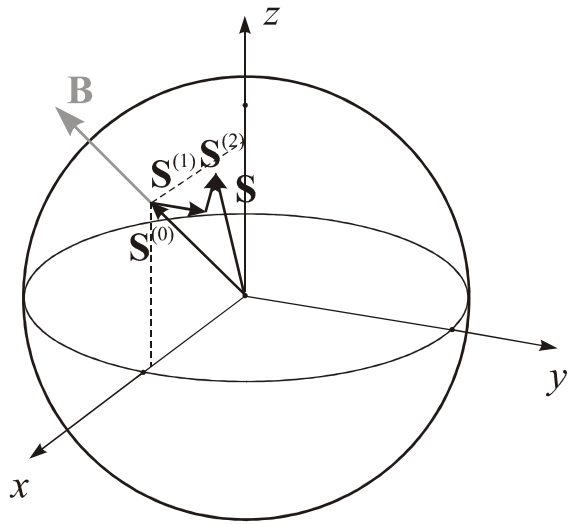

**Fig. 1.** Directions of the post-adiabatic corrections in the quasi-stationary solution on the classical-spin sphere. The case with $\dot{\theta} < 0$ and $\ddot{\theta} > 0$ is presented.

### 4. COMPARISON OF DIFFERENT APPROACHES AND GEOMETRIC PHASE

Now let us compare the obtained results to these from other approaches listed in the Introduction. In this way, we will pay particular attention to the phase of the wave function, Eq. (9).

First, Berry's approach [6] differentiates contributions to the phase from the geometry of spin evolution, Eq. (24) and those due to corrections to the frequency. In our case there is no correction due to geometry, and the second term of equation (45) in [6] brings about the same correction $\varphi^{(2)}$ as in our Eq. (9). In order to establish that, one should make a substitution $\theta \to \pi/2$, $\phi \to \theta$, $\tau \to \mathcal{B}\varepsilon t$ in Eqs. (42) and (45) of [6].

Second, the exact non-adiabatic equation for the phase given in [7,8] is of the same form as the adiabatic Berry phase expression, but on the classical-spin sphere. This leads to the substitution $\mathcal{B}/\mathcal{B} \to \mathbf{S}$ in the known expression for the Berry phase [3], so that

$$\varphi = \varphi^{(0)} - \frac{1}{2}\int_{\tilde{L}}\left(1 - \cos\tilde{\theta}\right)d\tilde{\phi}, \tag{26}$$

where $\tilde{\theta}$ and $\tilde{\phi}$ are the spherical coordinates of the qusi-stationary solution for $\mathbf{S}$, whereas $\tilde{L}$ is the contour of its evolution on the sphere. According to the solution (23), the coordinates of $\mathbf{S}$ are related to the coordinates of $\mathcal{B}/\mathcal{B}$, $\theta$ and $\phi$, in the approximation considered, as

$$\tilde{\theta} = \theta - \gamma, \quad \tilde{\phi} = \phi - \delta/\sin\theta. \tag{27}$$

Substituting Eq. (27) in Eq. (26), we arrive at

$$\varphi = \varphi^{(0)} + \varphi^{(1)} + \varphi^{(2)},$$

$$\varphi^{(1)} = \frac{1}{2}\int_L \left(1-\cos\theta\right)d\phi, \quad \varphi^{(2)} = -\frac{1}{2}\int_L \gamma\sin\theta\, d\phi + \frac{1}{2}\int_L \left(1-\cos\theta\right)d\left(\frac{\delta}{\sin\theta}\right). \tag{28}$$

Here $\theta$ and $\phi$ are the spherical coordinates of $\mathcal{B}/\mathcal{B}$, and $L$ is the contour of its evolution on the sphere. The phase $\varphi^{(1)}$ in Eq. (28) is the adiabatic Berry phase; in our case it vanishes since $\phi \equiv 0$. The first, related to the acceleration, term in $\varphi^{(2)}$ was overlooked in the example of paper [6] since only one rotation transformation was applied therein; it is also vanishes in our instance $\phi \equiv 0$. The second term in $\varphi^{(2)}$, Eq. (28), can be evaluated by integration by parts; it takes the following form (assuming that the integrated term vanishes):

$$\frac{1}{2}\int_L \left(1-\cos\theta\right)d\left(\frac{\delta}{\sin\theta}\right) = -\frac{1}{2}\int_L \delta\, d\theta = -\frac{1}{4}\int_0^t \frac{\dot{\theta}^2}{\mathcal{B}}dt, \tag{29}$$

which is exactly $\varphi^{(2)}$ in our equation (9).

Thus, we have shown that the phase $\varphi^{(2)}$ due to the second-order energy correction in (9), considered by Berry as purely dynamical [6], is a part of the non-adiabatic geometric phase (26) suggested in [7,8]. The second-order difference between the exact and adiabatic geometric phases has also been calculated in [11] for the case of a uniformly rotating magnetic field. However, the result there is twice as large as compared to the correction $\varphi^{(2)}$. (In order to establish that, one should make a substitution $\theta \to \pi/2$, $\omega_0 \to \mathcal{B}$, and $\omega \to \dot{\theta}$ in Eq. (18) of [11], and take into account that the first term in our $\varphi^{(2)}$, Eq. (28), vanishes for uniformly rotating field, $\gamma = 0$.) Apparently, this is a result of an arithmetic inaccuracy in [11], since the initial expressions for the exact and adiabatic phases there are equivalent to ours, given in Eqs. (26) and (28).

Finally, we consider the phase $\varphi^{(2)}$, Eq. (9), from the viewpoint of the generalized geometric phase approach put forward in [9,10]. Let us suppose that $\dot{\mathcal{B}} = 0$ for simplicity and introduce 2-dimensional generalized space of parameters: $\vec{M} = (\theta, \dot{\theta})$ (in the generic case it should be 4-dimensional with coordinates $\mathcal{B}$ and $\dot{\mathcal{B}}$ in addition); all vectors in this space will be marked by arrows. Then $\varphi^{(2)}$ can be presented as a contour integral in the generalized parameter space:

$$\varphi^{(2)} = -\int_l \frac{\dot{\theta}}{4\mathcal{B}}d\theta = -\int_l \vec{A}d\vec{M}, \quad \text{where} \quad \vec{A} = \left(\frac{\dot{\theta}}{4\mathcal{B}},\ 0\right) \tag{30}$$

is the vector potential on the $\vec{M}$-space and $l$ is the contour of the evolution in $\vec{M}$-space. The following non-zero field corresponds to this potential:

$$F = \frac{\partial}{\partial \vec{M}} \wedge \vec{A} = -\frac{1}{4\mathcal{B}}. \qquad (31)$$

Thus, for closed contours in the $\vec{M}$-space the generalized geometric phase can be presented as a surface integral of field $F$:

$$\varphi^{(2)} = -\oint_l \vec{A} d\vec{M} = -\int_s F \, d\vec{M} \wedge d\vec{M} = \frac{s}{4\mathcal{B}}, \qquad (32)$$

where $s$ − is the oriented area enclosed by the loop $l$ in $\vec{M}$-space (cf. [9,10]). Thus, in view of equations (30)–(32), the phase $\varphi^{(2)}$ is also the geometric phase on the generalized space $\vec{M}$. The fact that field $F$ is a non-zero one results from the non-local character of the phase $\varphi^{(2)}$.

## 5. SUMMARY

We have examined the spin evolution in a time-dependent in-plane magnetic field in the second-order adiabatic approximation. Both spinor and spin-sphere solutions have been obtained. While the spinor representation is useful for the phase derivation, it is more convenient to analyze the dynamics of the classical spin vector on the sphere. We have shown that the first- and second-order deflections of the spin precession axis appear due to the Coriolis forces. The second-order correction can also be addressed to the usual inertia due to non-uniform magnetic field rotation. The frequency of the precession acquires a second-order correction related to the shift of eigenvalues in the quantum problem.

We have considered the phase corrections and compared the results following from different approaches of [6–9,11]. It is shown that all the approaches (apart from an inaccuracy in [11]) lead to the same second-order correction in phase. Furthermore, a purely dynamical correction associated with the shift of eigenfrequencies [6] is, at the same time, a part of the exact geometric phase proposed in [7,8]. Remarkably, by using this approach, one can derive the quantum phase from the purely classical motion of the spin vector. The discussed phase correction can also be represented as a geometric phase on the generalized parameter space including parameters' derivatives [9].

Finally the quasi-stationary spin-sphere solution and the phase were given in the form, valid for the general case of a 3-dimensional magnetic field evolution, Eqs. (24) and (28)–(29).

## ACKNOWLEDGMENTS

I am grateful to S. Stenholm for fruitful discussions and indebted to Referees for pertinent remarks and helpful references. This work was partially supported by STCU (grant P-307) and CRDF (grant UAM2-1672-KK-06).

**REFERENCES**

1. F. Bloch, *Phys. Rev.* **70**, 460 (1946).
2. H. Klar and U. Fano, *Phys. Rev. Lett.* **37**, 1132 (1976); H. Klar, *Phys. Rev. A* **15**, 1452 (1977).
3. M.V. Berry, *Proc. R. Soc. A* **392**, 45 (1984); A. Shapere and F. Wilczek (eds), *Geometric Phases in Physics* (Singapore: World Scientific, 1989).
4. J.H. Hannay, *J. Phys. A: Math. Gen.* **18**, 221-230 (1985); M.V. Berry, *ibid.* **18**, 15 (1985).
5. K.P. Marzlin and B.C. Sanders, *Phys. Rev. Lett.* **93**, 160408 (2004), D.M. Tong *et al.*, *ibid.* **95**, 110407 (2005); A.K. Pati and A.K. Rajagopal, quant-ph/0405129; Z. Wu and H. Yang, *Phys. Rev. A* **72**, 012114 (2005); R. MacKenzie, E. Marcotte, and H. Paquette, *ibid.*, **73**, 042104 (2006); T. Vértesi and R. Englman, *Phys. Lett. A* **353**, 11 (2006).
6. M.V. Berry, *Proc. R. Soc. A* **414**, 31 (1987).
7. Y. Aharonov and J. Anandan, *Phys. Rev. Lett.* **58**, 1593 (1987); see also S.I. Vinnitski *et al.*, *Usp. Fiz. Nauk* **160**(6), 1 (1990) [*Sov. Phys. Usp.* **33**, 403 (1990)].
8. J.M. Souriau, *Structure des systemes dynamiques* (Paris, Dunod, 1970); *Structure of dynamical systems: a symplectic view of physics* (Boston, Birkhauser, 1997); P. Horváthy, *J. Math. Phys.* **20**, 49 (1979); J.R. Klauder, *Phys. Rev. D* **19**, 2349 (1979); H. Kuratsuji and S. Iida, *Phys. Rev. Lett.* **56**, 1003 (1986); H. Kuratsuji, *Phys. Rev. Lett.* **61**, 1687 (1988); J. Bolte and S. Keppeler, *Ann. Phys. (N.Y.)* **274**, 125 (1999).
9. K.Yu. Bliokh, *J. Math. Phys.* **43**, 5624 (2002).
10. K.Yu. Bliokh, *J. Phys. A: Math. Gen.* **36**, 1705 (2003); K.Yu. Bliokh and Yu.P. Stepanovskiy, *Zh. Eksp. Teor. Fiz.* **124**, 529 (2003) [*JETP* **97**, 479 (2003)].
11. D.M. Tong *et al.*, *Phys. Lett. A* **339**, 288 (2005).
12. S. Stenholm, *Laser Physics* **15**, 1421 (2005).